# Deflection of Ultra Slow Light by Earth's Gravity on Laboratory Length Scale

N. Kumar

*Raman Research Institute, Bangalore 560080, India*

Abstract

The high speed of light *in vacuo* together with the weakness of Earth's gravity rules out any experimental detection of gravitational deflection of light on the laboratory length scale. Recent advances in coherent optics that produce ultra slow light in highly dispersive media with the group velocities down to ~$10^2$ $ms^{-1}$, or even less, however, open up this possibility. In this work, we present a theoretical study for a possible laboratory observation of the deflection of such an ultra slow light in the highly dispersive medium under Earth's gravity. Our general relativistic calculation is based on the Gordon *optical* metric modified so as to include dispersion. The calculated linear vertical deflection turns out to be ~0.1 *mm* for a horizontal traversal of 0.1 *m*, and a group speed $v_g$ ~ $10^2$ $ms^{-1}$. Experimental realizability and some conceptual points involved will be briefly discussed.

Gravitational deflection of light by matter is one of the defining predictions of Einstein's General Relativity (GR) [1,2]. The extremely high speed of light *in vacuo* together with the extreme weakness of the gravitational coupling, however, requires stellar masses and astronomical length scales to give a sensible deflection. (Hence the eventful observation of the deflection of star light passing by the Sun during the total solar eclipse of 1919) [1]. But, the recent experimental demonstrations of ultra slow light [3-9] in highly dispersive transparent dielectrics, made possible by the advances in coherent optics, do offer some hope for observing the gravitational deflection of light under Earth's own gravity on a laboratory scale, *i.e.*, in a table-top experiment. Indeed, ultra slow light with the group velocity magnitude $v_g$ down to $10^2$ $ms^{-1}$ has been achieved through, *e.g.*, the Electromagnetic Induced Transparency (EIT) [4]. Here a sharp absorption dip is created within the spectral width of an atomic absorption line through an interference of the atomic transition amplitudes *via* alternative routes much as in the Fano resonance. The associated steepness of the frequency $(w)$ dependence of the real part of the refractive index $n(w)$ then leads, *via* the Kramers-Kronig (KK) dispersion relation, to a large group refractive index $n_g = n(w) + w\frac{dn(w)}{dw}$, and hence to a low group velocity, $v_g/c = 1/n_g \ll 1$. Encouraged by these developments, we have calculated the magnitude of the deflection to be expected from such a laboratory experiment. In fact, for $v_g \sim 10^2$ $ms^{-1}$, the calculated vertical deflection turns out to be appreciable $\sim 0.1$ *mm* for a horizontal traversal of 0.1 *m*. Below we give details of our GR calculation and discuss the results obtained.

It is a well known result central to General Relativity that a light ray follows the null geodesic of the given background space-time metric [2]. This, however, is so only in vacuum. In a medium, such as condensed transparent matter, the optical refractive index of the medium causes deviation of the light ray from the geodesic path. For a non-dispersive medium, it has been shown that the geometric optical ray again follows a null geodesic, but now it is the null geodesic of the so-called *optical* metric, due originally to Gordon [10,11] and used extensively ever since. Our theory is based on the *optical metric*, but with a modification that incorporates the physical effect of dispersion.

The optical metric $\tilde{g}_{\mu\nu}$ is related to the background gravitational metric $g_{\mu\nu}$ as [11]

$$\tilde{g}_{\mu\nu} = g_{\mu\nu} + (1 - \frac{1}{n^2}) U_\mu U_\nu \qquad (1)$$

where $U^\mu$ is the four-velocity of the optical medium in a state of general motion with $U_\mu U^\mu = -1$, and the background metric $g_{\mu\nu}$ has been used to lower the index, i.e., $U_\mu = g_{\mu\nu} U^\nu$. For the background metric $g_{\mu\nu}$ we adopt the Schwarzschild interval as appropriate to the spherical Earth [2]

$$ds^2 = g_{\mu\nu} dx^\mu dx^\nu = g_{tt} dt^2 + g_{rr} dr^2 + g_{\theta\theta} d\theta^2 + g_{\phi\phi} d\phi^2 \qquad (2)$$

with $x^0 \equiv t$, $x^1 \equiv r$, $x^2 \equiv \theta$, and $x^3 \equiv \phi$, and

$$g_{tt} = -\left(\frac{1}{1 - \frac{2GM_\oplus}{r}}\right), \quad g_{rr} = \left(1 - \frac{2GM_\oplus}{r}\right), \quad g_{\theta\theta} = r^2, \text{ and } g_{\phi\phi} = r^2 \sin^2\phi.$$

Here, $M_\oplus$ = mass of the Earth, $2GM = R_{\oplus G}$ the Earth's Schwarzschild radius, and we have set the speed of light *in vacuo* c = 1. Thus, the optical metric for a non-moving medium, as is the case considered now, becomes:

$$\tilde{g}_{tt} := \frac{1}{n^2} g_{tt}, \tag{3}$$

while the remaining components retain their Schwarzschild values.

Now, in the refractive medium the geometric optical ray follows the null geodesic of the optical metric given by

$$\frac{d^2 x^\alpha}{d\lambda^2} + \tilde{\Gamma}^\alpha_{\mu\nu} \left(\frac{dx^\mu}{d\lambda}\right)\left(\frac{dx^\nu}{d\lambda}\right) \tag{4}$$

with $\lambda$ the invariant parameter, and the optical connection (the Christoffel symbol) for the diagonal metric becomes

$$\tilde{\Gamma}^\alpha_{\mu\nu} = \frac{1}{2}\frac{1}{g_{\alpha\alpha}}\left(\frac{\partial g_{\alpha\mu}}{\partial x^\nu} + \frac{\partial \tilde{g}_{\alpha\nu}}{\partial x^\mu} - \frac{\partial \tilde{g}^{\mu\nu}}{\partial x^\alpha}\right)$$

with the optical metric coefficients given by Eq. (3).

The equations of motion for the null geodesic, confined in the $\phi$ = constant (= 0, say) plane, for the light ray now turn out to be

$$\frac{d^2 t}{d\lambda^2} = -\frac{2GM_\oplus}{r^2\left(1-\frac{2GM}{r}\right)}\left(\frac{dr}{d\lambda}\right)\left(\frac{dt}{d\lambda}\right)$$

$$\frac{d^2 r}{d\lambda^2} = -\frac{1}{n^2}\frac{GM}{r^2}\left(1-\frac{2GM}{r}\right)\left(\frac{dt}{d\lambda}\right)^2 - \frac{GM}{r^2\left(1-\frac{2GM}{r}\right)}\left(\frac{dr}{d\lambda}\right)^2 - r\left(1-\frac{2GM}{r}\right)\left(\frac{d\theta}{d\lambda}\right)^2,$$

$$\frac{d^2\theta}{d\lambda^2} = -\frac{2}{r}\left(\frac{d\theta}{d\lambda}\right)\left(\frac{dr}{d\lambda}\right), \qquad \ldots(5)$$

where the invariant (affine) parameter $\lambda$ is yet to be fixed. Note that the refrative index *n* enters only in the equation for the radial coordinate "*r*".

Given that the laboratory length scales are << geometrical radius of the gravitating Earth ($R_\oplus$) and $R_\oplus \gg R_G$ (the gravitational or the Schwarzschild radius of the Earth), we can now linearise the geodesic equations (4) as follows. Introduce the *horizontal* and the *vertical* displacements $\Delta x$ and $\Delta y$ through

$$r = \left[(R_\oplus - y)^2 + x^2\right]^{1/2} \approx R_\oplus - y$$
$$r_\theta \approx R_\oplus \theta = \Delta x \qquad (6)$$

with $r = R_\oplus$, $\theta = 0$ taken as the origin while the geodesic is confined to the $\phi = 0$ plane. It is assumed here that the light ray is launched horizontally from the origin. With these conditions, the linearized geodesic equations reduce to

$$\frac{d^2 t'}{d\lambda'^2} = \frac{2GM_\oplus}{R_\oplus^2} \frac{d\Delta y}{d\lambda'} \frac{dt'}{d\lambda'}$$

$$\frac{d^2 \Delta y}{d\lambda'^2} = \frac{GM_\oplus}{R_\oplus^2} \left(\frac{dt'}{d\lambda'}\right)^2$$

$$\frac{d^2 \Delta x}{d\lambda'^2} = 0, \qquad (7)$$

where we have scaled $\lambda$ and *t* as $\lambda' = \lambda / n$ and $t' = t / n$. The linearized equations are readily solved to give to the leading order,

$$\Delta y = \frac{1}{2} \left(\frac{GM_\oplus}{R_\oplus^2}\right) t'^2$$
$$\Delta x = V t' \qquad (8)$$

where *V* is a cosntant of integration, clearly related to the initial speed $(v = 1/n)$ of the optical ray launched horizontally from the origin. We must fix $\lambda$ now. In the present case of the small length scale limit, this is readily done by noting that the light ray must follow the optical null geodesic giving

$$ds^2 = 0 = -\frac{1}{n^2} dt^2 + R_\oplus^2 d\theta^2,$$ where we have approximated $1 - \frac{2GM}{R_\oplus} \approx 1$.

Thus $\Delta x \approx R_\oplus \Delta\theta = \left(\frac{\Delta t}{n}\right)$. This, combined with $\Delta x = V\lambda$, gives a physically consistent identification $V = \frac{1}{n}$ and $\lambda = \Delta x$. In order to traverse $\Delta x$, the time required will then be $n\Delta x$, giving

$$\Delta y = \frac{1}{2}\left(\frac{GM_\oplus}{R_\oplus^2}\right) n^2 \Delta x^2. \qquad (9)$$

The above derivation is for the non-dispersive medium. We now give a physically robust argument that for a dispersive medium all we have to do is to replace the refractive index *n* by the group refractive index $n_g$ that describes the wave-packet moving along the geodesic with the group speed $v_g = 1/n_g$. In order to see this, replace the monochromatic light wave of frequency $w$ by the superposition of a group of waves with a small frequency spread $\Delta w$. Inasmuch as the vertical displacement $\Delta y$ is of second order in $\Delta x$ (the horizontal displacement) with $\Delta y \ll \Delta x$, it is sufficient to consider the horizontal displacement only. Thus, all the wave components propagate along the optical null geodesic, but now the physically identifiable centre (the wave packet peak) will move with the group speed $1/n_g$, where $n_g = n + w\frac{dn}{dw}$. This follows from the very definition of the group velocity for a wave-packet [12]. Thus, we have the linear deflection $\Delta y$ in terms of the horizontal displacement $\Delta x$ as

$$\Delta y = \frac{1}{2}\left(\frac{GM_\oplus}{R_\oplus^2}\right) n_g^2 \Delta x^2, \qquad (10)$$

which is our final result.

Noting that $2GM/R_\oplus^2 \equiv$ the surface gravity of the Earth, the above expression for the linear deflections may be re-interpreted as the free fall of the light wave packet under gravity, with a post-Newtonian proviso that the equivalence principle of GR holds, namely that light too must fall. Indeed, the same is essentially true of the well known GR expression for the angular deflection $4GM/bc^2$ for the star light passing by the Sun, where 'b' is the impact parameter. The corresponding Newtonian expression for a test mass with velocity $v$ is $2GM/bv^2$. The two expressions are equal for $v=c$ to within a factor of 2. This asymptotic *time dilation* factor of 2 is, of course, absent in the present case of the short-length scale behaviour.

For a laboratory scale experiment with the ultra slow light $v_g \sim 10^2 \text{ ms}^{-1}$, and a horizontal traverse $\Delta x \sim 0.1 \text{ m}$, we obtain a vertical linear deflection $\Delta y \sim 0.1 \text{ mm}$ which is appreciable. It can be measured for a well collimated narrow laser beam using a position-sensitive detector. The deflection could be magnified through a multipass configuration. An interferometric detection should be much more sensitive. But, most importantly, the deflection is tunable – by tuning the group refractive index $n_g(w)$ over a narrow frequency range. One could adopt here a frequency modulation of $n_g(w)$ and the phase sensitive detection.

It is important that the dispersive optical medium be homogeneous so as not to force any non-gravitational deviation away from the geodesic. A fluid optical medium with EIT may be ideal. The above treatment can, of course, be readily generalized to the case of the optical fluid medium in general motion by taking $U^a \neq 0$.

In conclusion, therefore, we have derived an expression for the deflection of ultra slow light in a highly dispersive optical medium under Earth's gravity for a laboratory length scale experiment. The estimated deflection turns out to be measurable.

References

[1]   C.M. Will, *Theory and Experiment in Gravitational Physics* (CUP, Cambridge, 1993).
[2]   J.B. Hartle, *Gravity: An Introduction to Einstein's General Relativity* (Pearson Education, Singapore, 2003).
[3]   A. Casapi, M. Jain, G.Y. Yin and S.E. Harris, Phys. Rev. Lett., **74**, 2447 (1995).
[4]   L.V. Hau, S.E. Harris, Z. Dutton, and C.H. Behroozi, Nature **397**, 594 (1999).
[5]   M.M. Kash, V.A. Sautenkov, A.S. Zibrov, L. Holberg, R.G. Welch, M.D.Lukin, Y. Rostovtsev, E.S. Fry, and M.O. Scully, Phys. Rev. Lett. **82**, 5229 (1999).
[6]   M.S. Bigelow, N.N. Lepeshkin, and R.W. Boyd, Phys. Rev. Lett. **90**,113903 (2003); also Science **301**, 200 (2003).
[7]   P. Wu and D.V.G.L.N. Rao, Phys. Rev. Lett. **95**, 253601 (2005).
[8]   V.S. Zapasskii and G.G. Kozlov, Optics and Spectroscopy **100**, 419 (2006).
[9]   G. Piredda, J. Eur. Opt. Soc. (R) **2**, 07004 (2007)
[10]  W. Gordon, Ann. Phys. **72**, 421 (1923); Also, see Jurgen Ehlers, *Perspectives in Geometry and Relativity,* ed. B. Boffman (Indiana University Press, Bloomington, 1966), p. 127.
[11]  J.L. Anderson and E.A. Spiegel, Astrophys. J. **202**, 454 (1975).
[12]  M. Born and E. Wolf, *Principles of Optics* (Macmillan, Singapore, 1989).